\documentclass[10pt, conference, letterpaper]{IEEEtran}
\IEEEoverridecommandlockouts
\usepackage{cite}
\usepackage{amsmath,amssymb,amsfonts}
\usepackage{algorithmic}
\usepackage{graphicx}
\usepackage{textcomp}
\usepackage{xcolor}
\usepackage{amsthm,amssymb}
\usepackage{epstopdf}

\newcommand{\BS}{\Gamma_{\mathrm{BS}}}
\ifCLASSOPTIONcompsoc
\usepackage[caption=false,font=normalsize,labelfon
t=sf,textfont=sf]{subfig}
\else
\usepackage[caption=false,font=footnotesize]{subfig}
\fi
\usepackage[algoruled, vlined, linesnumbered]{algorithm2e}  
\usepackage{booktabs}
\DeclareMathOperator*{\argmax}{arg\,max}
\newtheorem{theorem}{Theorem}
\newtheorem{proposition}{Proposition}
\newtheorem{fact}{Fact}

\newcommand{\Cpx}{\mathbb{C}}
\newcommand{\ind}{1{\hskip -2.5 pt} \mathrm{I}}
\newtheorem{definition}{Definition}

\def\BibTeX{{\rm B\kern-.05em{\sc i\kern-.025em b}\kern-.08em
    T\kern-.1667em\lower.7ex\hbox{E}\kern-.125emX}}
\begin{document}

\title{Tracking the Best Beam for a Mobile User via Bayesian Optimization
}

\author{\IEEEauthorblockN{Lorenzo Maggi, Ryo Koblitz, Qiping Zhu, Matthew Andrews}
	\IEEEauthorblockA{\textit{Nokia} 
	}
}

\IEEEpubid{\makebox[\columnwidth]{\parbox{\columnwidth}{\copyright2023 IEEE. Personal use of this material is permitted. Permission
from IEEE must be obtained for all other uses, in any current or future
media, including reprinting/republishing this material for advertising or
promotional purposes, creating new collective works, for resale or
redistribution to servers or lists, or reuse of any copyrighted
component of this work in other works}\hfill}
\hspace{\columnsep}\makebox[\columnwidth]{ }}

\maketitle

\IEEEpubidadjcol

\begin{abstract}
  The standard beam management procedure in 5G requires the user
  equipment (UE) to periodically measure the received signal reference
  power (RSRP) on each of a set of beams proposed by the basestation
  (BS). It is prohibitively expensive to measure the RSRP on all beams
  and so the BS should propose a beamset that is large enough to allow a
  high-RSRP beam to be identified, but small enough to prevent excessive
  reporting overhead. Moreover, the beamset should evolve over time according to UE mobility.
  We address this fundamental performance/overhead trade-off via a
  Bayesian optimization technique that requires no or little
  training on historical data and is rooted on a low complexity
  algorithm for the beamset choice with theoretical guarantees.  We show
  the benefits of our approach on 3GPP compliant simulation scenarios.
\end{abstract}


\section{Introduction}
\label{sec:introduction}

Millimeter-wave (mmWave) frequencies are attractive for
next-generation wireless networks due to the large amount of bandwidth
available. One challenge with mmWave frequencies is the high
penetration loss but this can be mitigated by the gains achievable
with large antenna arrays \cite{pi2011introduction}. At high
frequencies, antenna spacing can be smaller thus more antennas can be
packed into a specified space, which in turn lets us use more
directive beams to compensate for the penetration
loss.  However, with a large antenna array it is prohibitively
expensive to have a separate radio chain controlling each antenna. A
common solution is the {\em hybrid beamforming} (HBF) architecture
\cite{heath2016overview} where analogue beams are created by
phase-shifters and digital precoding is performed on a set of radio
chains that is smaller than the total number of antennas.

In mmWave band 5G systems, the beam management is usually based on the
analogue beam domain and the system is designed to be compatible with
a hierarchical beam searching structure\cite{giordani2018tutorial}. In
initial access (IA), the synchronized signal block (SSB) can be
transmitted with wide beams while in data transmission, the channel
state information reference signal (CSI-RS) can be transmitted through
refined beams with the mainlobe contained within the selected wide
beam from IA.

The base station (BS) will sweep all the possible wide beams
periodically for IA, and it may also sweep all the possible refined
beams periodically for high refined beam gain tracking if it is
serving a large number of users. Each user equipment (UE) measures a
\emph{subset} of the beams and reports back to the BS the {\em Received
    Signal Reference Power} (RSRP) for each beam in the subset. From
these measurements the BS selects the ``best'' beam for the UE.  We
stress that this architecture does not
require any channel estimation, which becomes a challenge as the
number of antennas grows large.

After the analogue beam selection is complete, the BS then performs
scheduling, i.e., it chooses a set of UEs to receive data. 
Lastly,
digital precoding is performed by the radio chains to minimize the
interference across UEs.

In this work we focus on the selection of the analogue beams
for the downlink for which a specific UE must report RSRP measurements to the BS. Our scheme can be applied to both SSB and CSI-RS
beams and so in the sequel we do not make a distinction. The goal is to choose a beam for each UE that
maximizes the RSRP while limiting the UE reporting overhead, i.e., we
only want the UE to measure the RSRP on a small subset of the available beams before each selection.

\noindent\textbf{Scenario.} We assume a BS with $M$ antennas transmitting to a UE with $N$ antennas. Time is slotted and at slot $t$ the channel between the BS and the UE is represented by the matrix $H_t\in \Cpx^{N\times M}$.
A set $\BS\subset \Cpx^M$ of transmit beams is available to the BS
and a \emph{fixed} beam $u\in \Cpx^N$ is used by the UE\footnote{The UE usually has a small number of receiving antennas so the constructed receiving beam has a large beam main lobe. Hence, choosing a fixed received beam will not significantly affect the performance}.
If the BS selects beam $b_t\in \BS$ and transmits symbol $x_t\in \Cpx$ with power $\rho$ at time $t$, then the signal received by the UE is,
\begin{equation}
  y_t = \sqrt{\rho}u^*H_t b_t x_t + u^*n_t,
\end{equation}
where $\cdot^*$ and $n_t$ denote the conjugate transpose and the noise term, respectively. Then, we wish to select the beam $b_t\in \BS$ for data transmission during the slot $t$ so as to maximize the RSRP
$|y_t|^2$. If noise is circular Gaussian this amounts to
maximizing $\rho|\bar{H}_t b_t|^2$, where $\bar{H}_t=u^*H_t$ is the channel that the BS perceives, incorporating the UE's beam $u$.

Ideally, one would want to estimate channel $\bar{H}_t$
and find the element of $\BS$ that is closest to the
principal eigenvector of $\bar{H}_t^*\bar{H}_t$. However, the feedback required for channel estimation is prohibitive as the number of antennas increases. 
An alternative is for the BS to choose a set $B_t \subset
\BS$ of beams and ask the UE to measure the RSRP $|u^*H_tb|^2$ for all
$b\in B_t$ and report back each value. Then, the beam $b_t\in \BS$ with the highest RSRP is selected for data transmission by the UE during the current slot.
Yet, this procedure suffers from high feedback overhead if the UE has to measure a large number of beams. The goal of the {\em beamtracking} problem that we address in this paper is to select the beam $b_t$ for data transmission while trading off the achieved RSRP \emph{performance} with the {\em overhead}, i.e., how many beams the UE measures per slot. 

In this work we show how to choose beam $b_t$ via {\em Bayesian Optimization} (BO). The performance is evaluated according to 1) the {\em overhead} $|B_t|/|\BS|$, 2) the average RSRP {\em error} $|u^*H_tb_t|^2/\max_{b\in\BS}|u^*H_tb|^2$ and 3) the {\em accuracy}, i.e., the probability that $|u^*H_t b_t|^2=\max_{b\in \BS}|u^*H_t b|^2$. 



\subsection{Related work}

The beamtracking literature can be categorized into \textit{i)} 
\emph{RSRP-based}---which our contribution belongs to---where the BS determines
 the best beam for the UE only based on UE RSRP reports, \textit{ii)} 
 \emph{channel-based}, where the BS is assumed to be able to estimate the 
 channel or at least its covariance matrix, and \textit{iii)} 
 \emph{side-data assisted}, where additional information is required, such as 
 the GPS position of the UE.
Within the \textit{i)} \emph{RSRP-based} research thread, the work in \cite{kaya2021deep}, 
inspired by \cite{alahi2016social}, relies on the assumption that UE mobility 
tends to follow repeated patterns, and predicts the best beam for the next 
slots from previous RSRP measurements via a long short-term memory (LSTM) deep 
learning architecture. The main bottleneck is the training phase, during which 
the BS collects a large set of UE reports and trains an LSTM. References 
\cite{gottsch2020deep}, \cite{nishio2021wireless} predict the beam indexes 
with highest RSRP as well as blockage events, via deep learning. Similarly to our contribution, the work \cite{yang2022bayesian} uses Bayesian Optimization to estimate the best transmit and receive beams. However, the temporal aspect is not studied: once the user moves and/or the channel varies, the optimization has to be repeated from scratch.
The \textit{ii)} \emph{channel-based} thread is arguably the best investigated.
The contributions in \cite{va2016beam}, \cite{xin2019robust} rely on the assumption that the angles of arrival and departure of the channel evolve according to a Gauss-Markov model, and use a Kalman filter to track the main direction of the channel.
The work \cite{palacios2017tracking} exploits the ability of HBF transceivers to collect channel information from multiple spatial directions simultaneously, and designs two strategies (exhaustive beam search in a training phase and probabilistic beam tracking) to rapidly estimate the most suitable transmit/receive beams. For this scheme, the training effort is non negligible though. A sub-thread focuses on the assumption that, especially for mmWave, the channel has a sparse representation in the angular domain, i.e., only few scatterers exist.
This is exploited by estimating the channel via few linear measurements (which would in turn require the UE to report the complex received signal, instead of the RSRP) and then applying compressed sensing techniques, providing the main angles of arrival and departure of the channel, as in, e.g., \cite{chou2021wideband} and \cite{khordad2019kronecker}.
The \textit{iii)} \emph{side-data assisted} methods are more common in vehicular-to-infrastructure deployments, where the GPS location of the UE can be used for beamtracking, as in \cite{va2017inverse}, \cite{va2019online}, or via computer vision as in \cite{nishio2021wireless}.


\subsection{DFT Beam Construction} \label{sec:dft_beams}

The above problem was defined for an abstract beamset $\BS$, but
in practice $\BS$ typically consists of an array of
2-dimensional {\em Discrete Fourier Transform} (DFT) beams. In this
configuration the base station has a rectangular $M_H\times M_V$
antenna array with spacing $d_H,d_V$ in the horizontal and vertical
directions, respectively. The beamset $\BS$ is a
collection of DFT beams, defined on a grid of evenly
spaced azimuth angles $\{\theta_h\}_{h=1,\dots,H}$ and elevation angles $\{\phi_v\}_{v=1,\dots,V}$. The beam
$b_{h,v}$ is defined as the following $M_H\times M_V$ matrix:

{\small
\begin{align}
   & \frac{1}{\sqrt{M_H}}[1,e^{-j2\pi\frac{d_H}{\lambda}\sin\phi_v\cos\theta_h},\ldots,e^{-j2\pi\frac{d_H}{\lambda}(M_H-1)\sin\phi_v\cos\theta_h}]^T\otimes \notag \\
   & \ \frac{1}{\sqrt{M_V}}[1,e^{-j2\pi\frac{d_V}{\lambda}\cos\phi_v},\ldots,e^{-j2\pi\frac{d_V}{\lambda}(M_V-1)\cos\phi_v}],
\end{align}}
where $j=\sqrt{-1}$, $\otimes$ denotes the Kronecker product and $\lambda$
is the wavelength. If beam $b_{h,v}$ is selected then the
main lobe points in the direction with azimuth/elevation angles
$\theta_h,\phi_v$. 

\section{Bayesian optimization: Preliminaries} \label{sec:BOpreliminaries}

Bayesian optimization (BO) is a black-box optimization technique that
maximizes an unknown function $f:\mathcal X\rightarrow \mathbb R$,
where the domain $\mathcal X$ is a metric space, i.e., a (possibly
discrete) set endowed with a metric $\delta$. At each iteration $i$,
BO chooses a value of the input variable $x_i$, observes a (possibly)
noisy sample $\widetilde f(x_i)$ (also called the \emph{reward}) and
updates its estimate of $f$. BO is \emph{derivative-free} since it is
agnostic to the gradient of $f$ and does not attempt to estimate
it. BO is especially useful when a near-optimal point needs to be
found within a few iterations, due to the expense of evaluating $f$.
We refer to
\cite{shahriari2015taking} for an in-depth overview of BO. Next we
recall its salient features.

\noindent \textbf{Gaussian process.} In order to infer the value of the function $f$ at unseen points, BO relies on a statistical model that is typically a Gaussian Process (GP) \cite{williams2006gaussian}. Formally speaking, a GP is a collection of random variables, any finite collection of which has a multivariate Gaussian distribution. Hence, to define a GP we require a function defining the mean of each random variable and another function describing the covariance between any pair of variables.

The mean of the GP at each point $x$ is defined via the \emph{prior mean} function $m(.):\mathcal X\rightarrow \mathbb R$, which provides a reasonable estimation of $f(x)$, prior to any observation. By default, one can set $m(x)=\mathrm{constant}$ for all $x\in \mathcal X$. Yet, the choice of an informative prior by, e.g., domain knowledge and/or simulation helps BO to restrict the search region and avoid a cold start.

The covariance between any two elements of the GP is defined via a \emph{kernel} function $k(x,x'):\mathcal X \times \mathcal X \rightarrow \mathbb R$. Mat\'ern kernels \cite{williams2006gaussian} and the radial basis function (RBF) are classic examples of kernel functions. For instance, the RBF kernel is:
\begin{equation} \label{eq:rbf}
	k^{\mathrm{RBF}}_{\theta}(x,x') = \theta_1 \exp\left(\frac{\delta(x,x')}{\theta_2^2}\right), \quad \forall\, x,x'\in \mathcal X
\end{equation}
where $\delta$ is a distance metric and $\theta=[\theta_1,\theta_2]$ is the vector of hyper-parameters. The covariance between $f(x)$ and $f(x')$ is then computed as $\Sigma_{x,x'}= k_\theta (x,x') + \sigma^2 \ind(x=x')$, where $\sigma$ is the standard deviation of the observation noise and $\ind(.)$ is the indicator function.
The kernel determines the \emph{smoothness} of function $f$ with respect to the metric $\delta$. 

BO is an iterative process with three main
components. At each iteration we first {\em infer} the reward at
unmeasured points via the GP model. Then, we
pick a new point to measure. Finally, we tune the kernel
hyper-parameters.

\noindent \textbf{Inference.} Until iteration $i-1$ we have chosen points $\mathbf{x}_{i-1}=\{x_1,\dots,x_{i-1}\}$ and observed the corresponding rewards $\widetilde{\mathbf{f}}_{i-1}:=\{ \widetilde f(x_1),\dots,\widetilde f(x_{i-1})\}$. At iteration $i$ we want to infer the reward $f(x)$ for \emph{any} point $x$. By definition of a GP, the random variables $\widetilde{\mathbf{f}}_{i-1},f(x)$ are jointly Gaussian; moreover, their mean and covariance matrix can be obtained via the prior mean and kernel function, as described above. Therefore, we can infer $f(x)$ from previous measurements via the classic Gaussian posterior probability formula:
\begin{align}
	f(x)\, | \, \widetilde{\mathbf{f}}_{i-1},\mathbf{x}_{i-1} \sim  \mathcal N\Big( & \mu_{x} + \Sigma_{x,\mathbf{x}_{i-1}} \Sigma_{\mathbf{x}_{i-1}}^{-1}(\widetilde{\mathbf{r}}_i-\mu_{\mathbf{x}_{i-1}}), \notag      \\
	                                                                                & \Sigma_{x} - \Sigma_{x,\mathbf{x}_{i-1}} \Sigma_{\mathbf{x}_{i-1}}^{-1} \Sigma_{x,\mathbf{x}_{i-1}}^T \Big) \label{eq:posteriorGP}
\end{align}
where $\mu_{\mathbf{x}_{i-1}}=[m(x_1),\dots,m(x_{i-1})]$, $\Sigma_{x,\mathbf{x}_n}=[\Sigma_{x,x_n}]_{1\le n< i}$, and $\Sigma_{\mathbf{x}_i}=[\Sigma_{x_n,x_m}]_{1\le n,m< i}$.

\noindent \textbf{Choice of next point.} Choosing the next point $x_{i}$ is typically done by maximizing an \emph{acquisition function} $a$ that addresses the following \emph{exploration vs.\ exploitation} dilemma.
On the one hand, we want to exploit the learnings from previous observations and choose $x_{i}$ where the GP posterior mean is high.
On the other hand, to avoid getting stuck in local optima we should explore uncharted regions of $\mathcal X$ where the GP standard deviation is high.
A well-studied acquisition function is \emph{expected improvement} $a_{\mathrm{EI}}(x)$, computing the expectation of the improvement of the reward upon selecting $x$ with respect to the highest expected reward:
\begin{equation} \label{eq:ei}
	a_{\mathrm{EI}}(x)=\mathbb E \left[ f(x) - \max_{x'\in \mathcal X} \mathbb E[f(x')] \right]^+, \quad \forall\, x\in \mathcal X
\end{equation}
where expectations are with respect to the GP posterior.

Hyper-parameters $\theta,\sigma$ can be learned on-the-fly, by maximizing the log-likelihood of the collected reward samples.

The cumulative \emph{regret} of BO (with respect to the oracle solution that chooses the optimal point at all times) grows with the square root of the time horizon, as in \cite{KKS10}.

%

\section{Bayesian Optimization for Beamtracking}
\label{sec:BObeamtracking}

We now return to our beamtracking problem. We first present our beamset design principles, that we address via BO.

\subsection{Design principles} \label{sec:design_principles}

i) After a UE enters the cell, the BS wants to generate beamsets that can track the high RSRP beams in as few time slots as possible, since beams with low RSRP result in low data rate transmissions for the UE.

ii) The reason we can hope to do effective beamtracking without measuring all beams is that there are correlations in RSRP across the different beams in $\BS$ and across time. In particular, as the angular spread of the beams decreases, the RSRP function is increasingly smooth across $\BS$, and as the time slot frequency increases with respect to the channel coherence time, the RSRP function is smoother across time.

iii) The BS needs to determine how many beams should be proposed to the UE at each iteration; as the uncertainty on RSRP decreases, then fewer beams should be used. 


Next we show how points i)-iii) can be addressed via BO.

\subsection{Problem formulation via Bayesian optimization}

We model the unknown RSRP function for a UE $f_t(b):=|u^*H(t)b|^2$ as a
GP whose input variables are $x:=(t,b)$, where $t=0,1,\dots$
and $b\in\BS$. Here, time $t=0$ denotes the time that the UE
enters the cell served by the BS. As new RSRP measurements are
collected over time, we can infer the RSRP offered by a beam at the
next iteration via a GP surrogate model, analogous to \eqref{eq:posteriorGP}. In
particular, we can think of one BO iteration per time slot and so in
the sequel we shall use the terms ``iteration'' and ``time slot''
interchangeably.

There exist a few twists to the vanilla BO model introduced earlier. First, we 
have to deal with the
augmented time variable $t$, which we discuss in Section
\ref{sec:kernel_dft}. In particular, at a given time $t=t'$ we can
only request measurements of the form $f_{t'}(b)$. Second, we are not
simulataneously trying to approximate $f_t(b)$ for all $t,b$. At time
$t$ we are most interested in $f_{t'}(b)$ for $t'$ close to
$t$. Third, we have the freedom to choose \emph{multiple} beams
$B\subset \BS$ in each time slot, while the acquisition function
framework in vanilla BO only caters for a single function evaluation in each 
iteration.
Fourth, we do not simply want to maximize the performance of a beamset
in terms of RSRP, but we also wish to limit the associated beam
management reporting overhead. 

Next we describe our BO approach for beamtracking. We start by defining the
kernels, then we construct the prior mean for the GP. Finally, we show how to choose the beamset.

\subsection{Gaussian Process kernel design} \label{sec:kernel_dft}

The smoothness properties of $f$, discussed in point ii) above, are captured by the GP kernel $k(\cdot,\cdot)$. It is convenient \cite{richter2020model} to decouple the effect of beam and time variables and write the kernel as the product of two independent kernels:
\begin{equation}
	k_\theta \big((t,b),(t',b')\big) := k_{\theta}^{\mathrm{time}}(t,t') \times k_{\theta}^{\mathrm{beam}}(b,b').
\end{equation}

\subsubsection{Time kernel}

The time kernel $k_{\theta}^{\mathrm{time}}(t,t')$ describes the correlation of two RSRP measurements taken at iterations $t$ and $t'$. We want our beamtracking method to be applicable to any UE mobility pattern, which we do not even attempt to infer. The most robust choice is then to assume that $k_{\theta}^{\mathrm{time}}(t,t')$ fades as $|t-t'|$ increases; hence, the time kernel effectively decides the rate at which past samples are forgotten. Our choice for $k_{\theta}^{\mathrm{time}}$ is the RBF kernel (cf. Equation \ref{eq:rbf})
\begin{equation} \label{eq:rbf_iime}
	k^{\mathrm{time}}_{\theta}(t,t') := \theta_1 \exp\left(-\frac{t-t'}{\theta_2}\right)^2, \quad \forall\, t,t'\ge 0.
\end{equation}
where $1/\theta_2$ is the forgetting rate.

\subsubsection{Beam kernel for DFT beams}

To define the beam kernel $k_{\theta}^{\mathrm{beam}}(b,b')$ one has to first choose the metric $\delta$ describing the distance between two beams.
We propose here a natural approach based on the definition of DFT beams in Section~\ref{sec:dft_beams}. Since beams pointing in similar directions are expected to produce similar RSRP values, it is natural to define the kernel distance $\delta$ between two DFT beams $b_{h,v}$ and $b_{h',v'}$ as the weighted Euclidean distance between their indexes:
\begin{equation}
	\delta_{\ell}^{\mathrm{beam}}(b_{h,v},b_{h',v'}) = \sqrt{(h-h')^2/\ell_H + (v-v')^2/\ell_V}
\end{equation}
where the weights $\ell_H,\ell_V$ account for different spacing in azimuth and elevation of the DFT angle grid.
A classic kernel choice \cite{williams2006gaussian} is the Mat\'ern kernel $k_{\theta}^{\mathrm{beam}}(b,b')$, that writes:
\begin{align} \label{eq:matern}
	 & \frac{1}{\Gamma(\nu)2^{\nu-1}} \left( \sqrt{2\nu} \, \delta^{\mathrm{beam}}_\ell(b,b')\right)^\nu \kappa_\nu \left(\sqrt{2\nu} \, \delta^{\mathrm{beam}}_\ell(b,b') \right)
\end{align}
where the kernel hyper-parameters are $\theta=[\nu,\ell_H,\ell_V]$, $\kappa_\nu$ is the modified Bessel function of order $\nu$ and $\Gamma$ denotes the Gamma function. Importantly, the parameter $\nu$ controls the smoothness of the learned function.

\subsection{Gaussian process prior mean} \label{sec:prior_mean}

We also wish to make use of historical RSRP measurements to restrict the beam search for a new UE when it first connects to the BS. 
A natural way is to compute the GP prior mean $m_t(b):=m(b), \ \forall\,t$ as the average of the RSRP measurements reported by the UEs in the past to the same BS when beam $b\in \BS$ was deployed at the BS. This clearly gives a coarse estimation of $f$, but it can bias the beam search and rule out beams that never worked well in the past (e.g., beams with high elevation degree for the BS in rural areas with UEs located at low altitude).
Else, if historical data is not available at the BS, one can set $m_t(b)=\mathrm{constant}$ for all $t,b$.

\subsection{RSRP Inference}


At time $t$, the BS infers the function $f_t(b)$ for all $b\in \BS$ via the GP posterior formula \eqref{eq:posteriorGP}, where $x:=(t,b)$, with $t$ fixed and beam $b$ ranging over $\BS$, and where past sampling points are $\{(k,b')\}_{k=0,\dots,t-1,b'\in B_k}$.

\subsection{Beamset optimization via parallel acquisition function} \label{sec:beam_acq_fun_opt}

Next, we discuss how the BS chooses the next beamset $B_t$ on which the UE is asked to report RSRP measurements to the BS. First, we assume that the beam $b_t$ used by the UE for data transmission during slot $t$ is the one with highest RSRP among the proposed beamset $B_t$. We then define accordingly:
\begin{equation}
	f_t(B) := \max_{b\in B} f_t(b), \qquad \forall \, B\subset \BS, \ t= 0,1,\dots
\end{equation}
As in the classic BO framework, we choose the \emph{expected improvement} acquisition function (see Equation \ref{eq:ei}). However, to disincentivize sampling the entire beam dictionary we include a UE feedback overhead, modeled as a convex increasing function $h(.)$ of the beamset cardinality $|B|$, with $h(0)=0$. Then, the beamset $B_t$ chosen by the BS at time $t$ is given by:
\begin{equation} \label{eq:ei_beamtrack}
	B_t=\argmax_{B\subset \BS} \mathbb E \left[ f_t(B) - f^*_t \right]^+ - h(|B|),
\end{equation}
where $f^*_t$, as advocated in \cite{richter2020model}, is the highest RSRP that is \emph{believed} to be attainable across all beams at time slot $t$, i.e.,
\begin{equation}
	f^*_t := \max_{b\in \BS} \mathbb E[f_t(b)].
\end{equation}
The expression \eqref{eq:ei_beamtrack} is generally referred to as
\emph{parallel BO} \cite{frazier2018tutorial}, where multiple evaluations of
the unknown function are possible. To solve the combinatorial problem efficiently a number of approaches are available in the literature, but they either rely on
the assumption that observations are sequential \cite{ginsbourger2010kriging}
(while they occur at adjacent transmission units in our case, hence
practically simultaneously from a computation perspective) or that the GP domain $\mathcal X$ is continuous
\cite{wang2020parallel} (while $\mathcal X$ is inherently
discrete in our case).
Moreover, such approaches are particularly suited when the dimension of $\mathcal X$ is large (whereas it is just 2 in the case of DFT beams).
Therefore, in the next section we derive an efficient method tailored for our use case that approximates the optimal $B_t$ with low complexity and theoretical guarantees.

\subsubsection{Greedy algorithm with theoretical guarantees}

We now design a method that approximates the optimal $B_t$ with low complexity and theoretical guarantees.
Since our analysis holds for any iteration $t$, we will omit subscript $t$.

\noindent \textbf{Auxiliary problem.}
We focus first on a simplified version of problem \eqref{eq:ei_beamtrack}, where the beamset size is fixed and equal to $n$:
\begin{equation} \label{eq:ei_beamtrack1}
	J^*(n)=\max_{B\subset \BS:|B|=n} J(B) := \mathbb E \left[ f(B) - f^* \right]^+.
\end{equation}
We will prove that $J(B)$ is a \emph{monotone} and \emph{submodular} function of the beamset $B$. Monotonicity states that larger sets bring higher rewards. Submodularity is analogous to concavity and claims that the incremental reward of adding an element to a certain initial set decreases as the initial set enlarges.

\begin{definition}
	The set-valued function $J$ is \emph{monotone} if $J(\mathcal B)\le J(\mathcal B')$, for all $\mathcal B\subset \mathcal B'$.
\end{definition}

\begin{definition}
	The set-valued function $J$ is \emph{submodular} if, for all $\mathcal B \subset \mathcal B'$ and $ b^*\notin \mathcal B'$, $J(\mathcal B\cup b^*) - J(\mathcal B) \ge J(\mathcal B'\cup b^*) - J(\mathcal B')$.
\end{definition}

\begin{proposition} \label{thm:beamtrack_ihm}
	Function $J(.)$ is monotone and submodular.
\end{proposition}
\begin{IEEEproof}
	To prove monotonicity we have to show that
	\[
		\mathbb E \left[ f(B) - f^* \right]^+ \le \mathbb E \left[ f(B') - f^* \right]^+, \quad \mathrm{if} \ B\subset B'\subset \BS.
	\]
	This simply stems from the fact that $f(B)$ is the maximum over a set of random variables $\{f(b)\}_{b\in B'}$ that is strictly larger than $\{f(b)\}_{b\in B}$.
	To prove submodularity we observe that, if $B\subset B'$ and $b^*\notin B'$,
	\begin{align}
		[f(B\, \cup \, b^*)- & \,f^*]^+ - [f(B)-f^*]^+ \label{eq:subm1}                 \\
		=                    & \, [f(b^*)-\max\{f^*,f(b) \ \forall\,b\in B\}]^+         \\
		\ge                  & \, [f(b^*)-\max\{f^*,f(b) \ \forall\,b\in B'\}]^+        \\
		=                    & \, [f(B'\cup b^*)-f^*]^+-[f(B')-f^*]^+. \label{eq:subm2}
	\end{align}
	By taking the expectation of \eqref{eq:subm1},\eqref{eq:subm2} one obtains the submodularity definition of $J$, q.e.d.
\end{IEEEproof}

By exploiting a classic result in combinatorial analysis  \cite{nemhauser1978analysis} we can claim that a simple greedy algorithm that adds iteratively the beam maximizing the incremental expected improvement (Algorithm \ref{alg:beamtrack_greedy}) achieves an optimality gap of $e^{-1}$.

\begin{theorem} \cite{nemhauser1978analysis} \label{thm1}
	Let $\underline{J}=\min_{b\in \BS} J(\{b\})$. Let $J(B^{\mathrm g}(n))$ be the reward achieved by Algorithm \ref{alg:beamtrack_greedy}. Since $J$ is monotone and submodular, then the optimality gap is bounded by $e^{-1}$:
	\begin{equation}
		\frac{J^*(n)-J(B^{\mathrm g}(n))}{J^*(n)-\underline J} \le e^{-1} \approx 0.37, \quad \forall \, n\ge 1.
	\end{equation}
\end{theorem}


\begin{algorithm}{\small
	\textbf{Initialization:} Set $B^{\mathrm g}(0):=\emptyset$.\\
	\For{$k=1,\dots,n$}{
	Compute $b^{\mathrm g}(k):=\argmax_{b\in\BS \setminus B^{\mathrm g}(k-1)} J(B^{\mathrm g}(k-1) \cup b)$\\
	Set $B^{\mathrm g}(k) :=B^{\mathrm g}(k-1) \cup b^{\mathrm g}(k)$\\
	}
	\Return beamset $B^{\mathrm g}(n)$
	\caption{(Auxiliary) Fixed size beamset choice.}
	\label{alg:beamtrack_greedy}}
\end{algorithm}

\noindent \textbf{Optimized beamset.} We can now finally address our original problem in \eqref{eq:ei_beamtrack}, where the beamset size is \emph{not} fixed.
In principle, one could run Algorithm \ref{alg:beamtrack_greedy} for all $n$'s and then choose the beamset $B^{\mathrm g}(\bar{n})$ with highest objective $J(B^{\mathrm g}(\bar{n}))-h(\bar{n})$, for some $\bar n\ge 1$.
Yet, recomputing the optimized beamset $B^{\mathrm g}(n)$ from scratch for every $n$ is redundant: the iterative nature of greedy Algorithm \ref{alg:beamtrack_greedy} suggests that, once $B^{\mathrm g}(n)$ is computed, one only has to add $b^{\mathrm g}(n+1)$ to obtain $B^{\mathrm g}(n+1)$.
Moreover, it is not necessary to add beams indefinitely, but only until a limited size. To show this, we first observe that $J(B^{\mathrm{g}}(n))$ is the discrete version of a concave increasing function.

\begin{fact} \label{prop:decr_incr}
	The function $J(B^{\mathrm{g}}(n))$ is increasing in $n$, while its increments are decreasing in $n$, i.e., for all $n>1$, i.e., $J(B^{\mathrm{g}}(n+1))-J(B^{\mathrm{g}}(n)) \le J(B^{\mathrm{g}}(n))-J(B^{\mathrm{g}}(n-1))$.
\end{fact}


The difference between a concave increasing function ($J$) and a convex increasing function ($h$) has at most one inflection point. Thus, to approximate the beamset selection problem \eqref{eq:ei_beamtrack} it suffices to add beams iteratively as in greedy Algorithm \ref{alg:beamtrack_greedy}, until the objective function $J(B^{\mathrm g}(n))-h(n)$ starts decreasing.
We recap this procedure in Algorithm \ref{alg:beamtrack_greedy_any_n}, used by the BS to compute at each time slot $t$ the beamset $B_t$.

\begin{algorithm} {\small
	\textbf{Goal:} maximize expected improvement as in \eqref{eq:ei_beamtrack}.\\
	\textbf{Initialization:} Set $B^{\mathrm g}(1):=\argmax_{b\in\BS} J(b)$.\\
	\For{$n=2,3,\dots$}{
	Compute $b^{\mathrm g}(n)=\argmax_{b\in\BS \setminus B^{\mathrm g}(n-1)} J(B^{\mathrm g}(n-1) \cup b)$\\
	Set $B^{\mathrm g}(n) :=B^{\mathrm g}(n-1) \cup b^{\mathrm g}(n)$\\
	\If{$J(B^{\mathrm g}(n))-h(n)\le J(B^{\mathrm g}(n-1))-h(n-1)$}{\Return beamset $B_t=B^{\mathrm g}(n-1)$.}
	}
	BS proposes beamset $B_t$ to the UE at time slot $t$.
	\caption{Choice of beamset $B_t$ at time slot $t\ge 0$}
	\label{alg:beamtrack_greedy_any_n}}
\end{algorithm}

\noindent \textbf{Practical implementation of Algorithm \ref{alg:beamtrack_greedy_any_n}.} A closed formula for $J(B)$ is only known for $|B|=1,2$ (see \cite{williams2006gaussian}, \cite{ginsbourger2010kriging}, respectively). Hence, in practice, $J(B)$ should be estimated via Monte-Carlo sampling from the GP posterior distribution for $|B|\ge 3$. To further reduce complexity, efficient sampling methods with linear complexity in the number of observations can be used, such as the one in \cite{wilson2020efficiently}.\\

\begin{figure}[htb]
	\centering
	\includegraphics[width=0.5\textwidth]{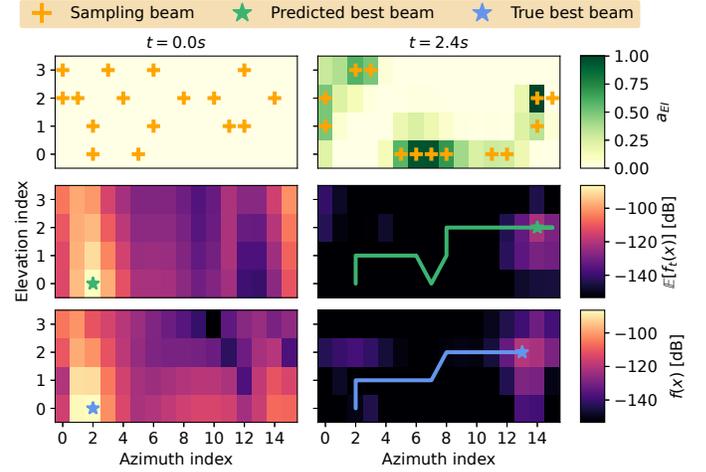}
	\caption{Time history of a typical UE at time slots $0, 30$. Top row: Acquisition function
		(Expected Improvement) used to choose sample beam indices (orange crosses). Center
		row: Posterior mean with predicted best beam index (green star) and path
		trace (green line) overlaid. Bottom row: True RSRP landscape with
		true best beam (blue star) and path trace (blue line) overlaid.
		\label{fig:colormaps}}
\end{figure}

\begin{algorithm}{\small 
	\SetAlgoLined
	\textbf{Initialization.}
	Choose the kernel function $k_\theta$ as the product of beam and time kernels as in Sect. \ref{sec:kernel_dft}\;
	BS computes the prior mean $m$ as in Sect. \ref{sec:prior_mean}\;
	BS initializes the hyper-parameters $\theta,\sigma$\;
	UE connects to BS at time $t=0$.\\
	\For{time slot $t=0,1,\dots$}{
		BS computes the beamset $B_t$ via Algorithm \ref{alg:beamtrack_greedy_any_n}\;
		UE reports the RSRP for each beam in  $B_t$\;
		BS uses beam $b_t\in B_t$ with highest reported RSRP for data transmission to the UE until next slot\;
		BS updates $\theta,\sigma$ via max-likelihood\;
	}
	\caption{BO for beamtracking}
	\label{algo:BOGP_beamtrack}}
\end{algorithm}

For the reader's convenience, in Algorithm \ref{algo:BOGP_beamtrack} we recap
the main steps of our BO algorithm for beamtracking. In Figure \ref{fig:colormaps} we provide a typical UE time history visualizing the 2D acquisition function, posterior distribution, and ground truth.

\section{Numerical Results} \label{sec:num_results}

We evaluate our approach using a 5G NR 3GPP-compliant system level simulator and benchmark its performance against single-slot (spatial) algorithms that only use RSRP information from the current timeslot and an LSTM-based
multi-slot (spatio-temporal) algorithm that (like BO) also utilizes past measurements. In out BO setup we use a
non-informative prior mean function (i.e., $m_t(b)=0$ for all $t,b\in \BS$), mimicking execution at the BS without any
offline training or {\em a priori} tuning. The
simulation specifications are found in Table
\ref{tab:simparams}. 

\begin{table}[htb]
	\centering
	\caption{Simulation configuration parameters\label{tab:simparams}}
	\begin{tabular}{p{0.1\textwidth}p{0.3\textwidth}}
		\hline 
		Scenario                       & 3D-UMi-street Canyon                        \\
		Deployment                     & Hexagonal grid, 7 BS sites, 21 cells,               \\
		& BS antenna height 10m (downtilt $10^\circ$) \\
		ISD                            & 100m                                        \\
		Carrier                        & 28 GHz                                      \\
		Bandwidth               & 50 MHz                                      \\
		Spacing            & 120 kHz                                     \\
		Frame                & TDD, DL data frame only                     \\
		gNB antenna                    & (M, N, P, Mg, Ng) = (16, 16, 2, 1, 1)       \\
		& dual-polarized panel arrays                 \\
		gNB GoB & 64 Tx beams: Tx Beam Azimuth (deg)          \\
		& $=-56.25+7.5n$, $n=0,\ldots,15$             \\
		& Elevation (deg) $= \{0,7.5,15,22.5\}$       \\
		UE antenna       & (M, N, P) = (2, 2, 2) per dual-polarized panel            \\
		& UE orientation uniformly distributed        \\
		HARQ                           & No retransmission                           \\
		Traffic Models                 & Traffic model: full buffer                  \\
		UE distribution                & 10 UEs/sector, randomly distributed         \\
		& 100\% of UEs outdoor                        \\
		UE speed                       & {30, 45, 60, 75, 90} km/h                   \\
		Time slot                      & 80ms                                        \\
		UE trajectory            & Straight with random direction              \\
		Channel    & 3GPP spatial consistency with proc.\ A \\ 
		\hline 
	\end{tabular}
\end{table}

\begin{table*}
	\centering
	\caption{Performance of spatial and spatio-temporal methods at varying UE speeds.}
	\begin{tabular}{@{}l l c c c c c c c c c@{}}\toprule
		         &                    & \multicolumn{3}{c}{$s=30\,$km/h} & \multicolumn{3}{c}{$s=60\,$km/h} & \multicolumn{3}{c}{$s=90\,$km/h}                                                             \\\cmidrule(lr){3-5}\cmidrule(lr){6-8}\cmidrule(lr){9-11}
		         &                    & Accuracy                         & Overhead                         & RSRP error                            & Accuracy & Overhead &
		RSRP error    & Accuracy           & Overhead                         & RSRP error                                                                                                                           \\
		\midrule
		Spline   & $\phi=0.25$        & 0.804                            & 0.25                             & 2.36                             & 0.781    & 0.25     & 2.35    & 0.770   & 0.25  & 2.37    \\
		         & $\phi=0.5$         & 0.934                            & 0.5                              & 0.653                            & 0.926    & 0.5      & {0.639} & {0.916} & 0.5   & {0.669} \\
		GPR      & $\phi=0.25$        & 0.329                            & 0.25                             & 13.1                             & 0.353    & 0.25     & 12.0    & 0.361   & 0.25  & 11.3    \\
		         & $\phi=0.5$         & 0.885                            & 0.5                              & 1.21                             & 0.880    & 0.5      & 1.14    & 0.872   & 0.5   & 1.11    \\
		PredRNN  & $(J, K) = (5,5)$   & 0.933                            & 0.5                              & 0.217                            & 0.924    & 0.5      & 0.294   &
		0.903    & 0.5                & 0.507                                                                                                                                                              \\
		         & $(J, K) = (5, 15)$ & 0.891                            & 0.25                             & 0.465                            & 0.873    & 0.25     &
		0.705    & 0.815              & 0.25                             & 1.66                                                                                                                            \\\midrule
		BayesOpt & low overhead       & 0.943                            & 0.116                            & 0.627                            & 0.900    & 0.122    & 1.05    & 0.874   & 0.126 & 1.23    \\
		         & high accuracy      & 0.961                            & 0.195                            & 0.425                            & 0.931    & 0.207    & 0.700   & 0.908   & 0.21  & 0.929   \\
		\bottomrule
	\end{tabular}
	\label{tab:results}
\end{table*}

We first compare our method to two spatial interpolation approaches: a
Scipy implementation of (rectilinear bivariate) spline interpolation~\cite{2020SciPy-NMeth}, and a
Scikit-Learn
implementation of Gaussian process regression~\cite{scikit-learn} with a Mat\'ern 3/2 kernel.
For these interpolation methods, we first sample a subset of beams
$B_{\phi}\subset\Gamma_\mathrm{BS}$ where the subscript $\phi$ denotes the
proportion of $\Gamma_\mathrm{BS}$, and choose the best beam $b_t$ according to
$\argmax_{b\in B_\phi}\mathrm{RSRP}(b)$. We consider two sampling fractions:
$\phi\in\{0.25, 0.5\}$. For our BO method, we consider two setups: a \emph{high accuracy} configuration
prioritizing accuracy by allowing for a larger sampling beamset cardinality
and encouraging more exploration by reducing the overhead penalty function $h(.)$;
a \emph{low overhead} configuration prioritizing overhead by
restricting the beamset cardinality to be at most 16 (i.e.\ $12.5\%$ of the 64 available beams in $\BS$) and increasing the penalty function $h(.)$, thereby making the sampling more greedy.

The performance of all schemes for 3 different UE speeds is shown in
Table~\ref{tab:results}. Here, RSRP error is measured in dB. Gaussian
process regression (GPR) requires a sampling
fraction of 0.5 to achieve acceptable accuracy and RSRP error,
although this error is above 1 dB across all speeds even with
a sampling fraction of 0.5. Of these two interpolation methods, spline interpolation performs the best across all three metrics,
which is notable since it has a lower computational
complexity.

The high accuracy BO method achieves over 90\% accuracy and sub 1 dB RSRP error with an
overhead around 20\% across all speeds tested. Only spline interpolation achieves
marginally higher accuracy and lower RSRP error at a UE speed of 90
km/h, albeit with over double the overhead. Even with an overhead of only 12\%,
the low overhead BO method is able to achieve approximately 90\% accuracy and 1 dB RSRP
error at UE speeds lower than 60 km/h.

Finally, we compare our BO method for beamtracking to a spatio-temporal algorithm described in \cite{NokiaTdoc} that utilizes a reimplementation of
PredRNN, a recurrent neural network for predictive learning using
Long Short-Term Memory (LSTM) units \cite{Wang2017LSTM}.
PredRNN uses a so-called unified memory pool, allowing the spatio-temporal LSTM
units to extract both spatial and temporal representations simultaneously. We refer to \cite{Wang2017LSTM} for full details. PredRNN \cite{NokiaTdoc} predicts the \emph{most probable}
length-$K$ sequence of best beams $b^\ast$ given the previous length-$J$
sequence including the current observation:
\[
	b^\ast_{t+1}, \dots,b^\ast_{t+K} =
	\argmax_{b_{t+1},\dots,b_{t+K}}
	p \left(b_{t+1}, \dots b_{t+K}|b^\ast_{t-J+1},\dots b^\ast_t\right).
\]
Given that PredRNN requires noiseless inputs, we sample {\em all} beam indexes
for $J$ time slots and then predict the next $K$ time slots, repeating this
sampling/prediction cycle until
the UE has left the domain. In Table~\ref{tab:results} we show performance for
$(J,K)=(5,5)$ (i.e., $\phi=0.5$) and for $(J,K)=(5,15)$ (i.e., $\phi=0.25$).
PredRNN typically achieves high accuracy and low RSRP error when the overhead is 0.5 (which is higher overhead than for the BO methods),
but the accuracy and RSRP error deteriorate as we lower the overhead to 0.25.

\begin{figure}
	\centering
	\includegraphics[width=\linewidth]{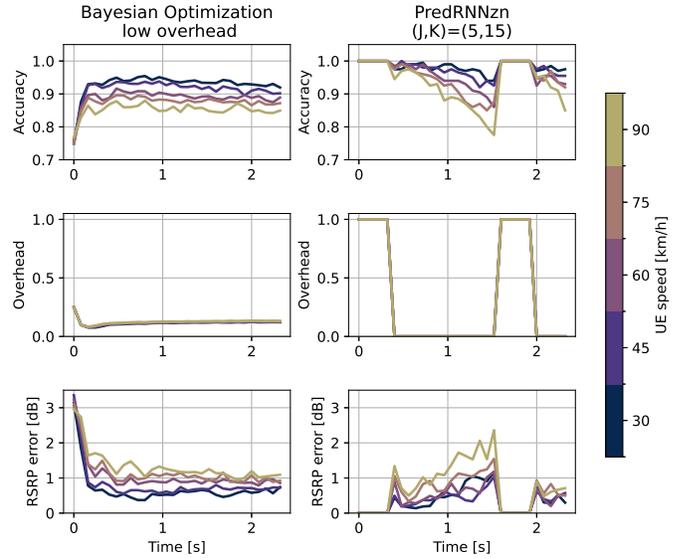}
	\caption{Left column: Bayesian optimization (low overhead configuration)
		evolution of accuracy, overhead, and RSRP error for varying UE
		speeds. Right column: PredRNN ($(J,K) = (5, 15)$ configuration) evolution of
		accuracy, overhead, and
		\emph{RSRP} error for varying UE speeds.\label{fig:convergence}}
\end{figure}

Figure~\ref{fig:convergence} shows the evolution
of accuracy, overhead, and RSRP error for both BO and PredRNN (left and right columns, respectively).
For our BO-based approach, accuracy and RSRP error degrade with
increasing UE speed, as seen in
Table~\ref{tab:results}.
Interestingly, the accuracy and RSRP error improve approximately monotonically,
while there is a distinct undershoot in the overhead metric near $t\approx
	0.3$ before converging towards approximately 12.5\% (or 20\% for the high
accuracy configuration). We attribute this undershoot to the development of the
time kernel during the cold-start: without \emph{a priori} knowledge the RSRP
landscape is initially treated as static (with a large temporal length scale), but as
RSRP measurements come in temporal correlations become apparent, driving the
temporal length scale down. The shift from static to dynamic RSRP landscape
manifests itself in a broadening of the posterior, promoting more
exploration and an increase in the overhead.

For PredRNN, the overhead is at
100\% during the training phase and then comes down to 0\% during
the prediction phase. The accuracy and RSRP error start to deteriorate as soon
as we enter the prediction phase, especially for the higher UE speeds. This is
in contrast to our BO-based method, which after the cold-start, displays
consistent performance over the remaining slots.
A major difference
between our BO-based method and PredRNN is that the latter requires
offline training whereas the BO results in this section are obtained
entirely online with \emph{no pretraining}--although in principle BO can leverage historical data to train its prior mean, as discussed in Section \ref{sec:prior_mean}.

\section{Conclusions}
\label{sec:conclusions}

In this paper we have described how Bayesian Optimization (BO) provides an
effective paradigm for beamtracking so that the UE can maintain
connection to a high-RSRP beam by measuring a limited set of beams in every
time slot. There are multiple ways in which this work can be
extended. First, if the beam dictionary is more exotic than classic DFT then the kernel choice is less evident. One could use \emph{meta-learning} techniques \cite{rothfuss2021pacoh} where the metric $\delta$ computing the distance between points in $\mathcal X$ is defined by a neural network.
Second, one could extend our approach by concurrently scheduling wide and narrow beams, to achieve increased flexibility and reduced overhead with respect to the standard hierarchical wide (SSB) and narrow (CSI-RS) beam selection in 5G.



\end{document}